\documentclass[12pt]{article}
\usepackage{epsfig}
\begin{document}
\vspace*{-.6in} \thispagestyle{empty}
\begin{flushright}
SUSX-TH/02-007

\end{flushright}
\baselineskip = 15pt

\vspace{.5in} {\Large
\begin{center}
{\bf Towards a Realistic Picture of CP Violation  in
 Heterotic String  Models  }

\end{center}}

\vspace{.5in}

\begin{center}

  Oleg Lebedev  and Stephen
Morris

\vspace{.5in}

\emph{Centre for Theoretical Physics, University of Sussex,\\
Falmer, Brighton, BN1 9QJ, U.K.}
\end{center}

\vspace{.5in}

\begin{abstract}
We find that dilaton dominated supersymmetry breaking 
and spontaneous CP violation
can be achieved  
in heterotic string  models with superpotentials singular
at the fixed points of the modular group.
A semi--realistic picture of CP violation emerges in such models:
the CKM phase appears due to a complex VEV of the $T$-modulus, while the soft
supersymmetric CP phases are absent due to an axionic--type symmetry.

\end{abstract}

\noindent

\newpage

\pagenumbering{arabic}

\section{Introduction}

While string theory remains an excellent candidate for the theory of everything, 
its connection to the presently observed world remains obscure. 
In this letter, we attempt to bridge one of the gaps, that is to 
address the problem  of CP violation. 
Recent observations have shown that CP is heavily violated in
the CKM (Cabibbo-Kobayashi-Maskawa) mixing matrix \cite{Aubert:2001nu}. 
On the other hand, if we are to retain low energy supersymmetry as a solution
to the hierarchy problem,
the electric dipole moment (EDM) experiments \cite{mercury}  require
the soft SUSY CP-phases to be vanishingly small (see \cite{Abel:2001vy} for a recent review).
Thus, the challenge is to find a supersymmetric string model
which produces a large CKM  phase while
having small enough soft SUSY CP-phases. The problem is exacerbated by the fact
that even if the soft CP-violating phases are absent initially, they are often
induced by a quark superfield basis rotation  \cite{Abel:2001cv}.

It is well known that CP is a gauge symmetry in string
theory \cite{Dine:1992ya} and therefore   must be broken spontaneously. 
Natural candidates for breaking CP are the dilaton ($S$) and 
moduli\footnote{There are three such moduli in most models, one
for each complex plane, but here we will assume that they all take the
same value.} ($T$) fields
\cite{Strominger:1985it}
which are common to string models. The former, however, cannot produce the CKM phase,
so a complex $\langle T \rangle$ is required in realistic models. Of course,
CP violation may originate from an entirely different sector, but this would be 
highly model--dependent and so we do not discuss this possibility here.
 
In what follows, we will concentrate on string models possessing target
space modular invariance, such as  heterotic  orbifolds.
That is, physics is invariant under the $PSL(2,Z)$ transformations

\begin{equation}\label{T-dual}
  T\longrightarrow\frac{a T-ib}{i c T+d}\;,
\end{equation}

\begin{equation}\label{strans}
  S \longrightarrow S+\frac{3}{4 \pi^2} \delta_{GS} \ln (i c
  T+d)  \;,\
\end{equation}

\noindent where $a,b,c,d$ are integers obeying $ad-bc=1$ and
$\delta_{GS}$ is the Green-Schwarz anomaly cancellation
coefficient. This symmetry imposes strict constraints
on the form of the effective superpotential and plays
a crucial role in our analysis. As in our earlier studies \cite{Khalil:2001dr},
we impose the following
phenomenological constraints: \\ \ 

1. the dilaton is stabilized at Re$S \sim 2$

2. the pattern of CP violation is phenomenologically acceptable

3. a realistic SUSY breaking scale \\ \ 

Previous attempts to produce CP violation \cite{Bailin:1998iz}
did not address  the problem of
dilaton stabilization and thus were not fully realistic.  
Additional constraints such as the absence of the flavor changing neutral currents (FCNC)
can also be imposed, but these are often safisfied automatically in this class
of models if a non-trivial CKM phase is produced at the renormalizable level (due to the 
Yukawa coupling selection rules) \cite{Khalil:2001dr}.

Dilaton stabilization has an important effect on the pattern of 
supersymmetry breaking. Our previous studies \cite{Khalil:2001dr}
have shown that it often forces moduli--dominated supersymmetry
breaking which has a disastrous phenomenology. On the other hand,
when dilaton--dominated SUSY breaking was produced, no CP violation
appeared. 

In the present letter, we will try to reconcile CP violation
and correct supersymmetry breaking by relaxing the assumption
that the superpotential has no singularities in the fundamental domain
of the modular group.  The singularities appear at the points in the moduli
space where the threshold corrections to the gauge couplings become
infinite. This, of course, happens at Re$T\rightarrow\infty$ corresponding
to a large contribution from light Kaluza-Klein states, but may also 
occur at other points in the moduli space  \cite{Cvetic:1991qm}.
Since explicit examples exhibiting this behavior are lacking,
we will take  the {\it bottom--up} approach, i.e. adopt the above
phenomenological requirements  as our strarting point
while being consistent with the modular invariance. 
We find that singular superpotentials allow for phenomenologically interesting
minima at which CP violation is present in the Standard Model sector  but
not in the soft SUSY breaking terms.

We shall proceed as follows. In the next section, we present our
framework. In section 3,  
we discuss patterns of the minima of the scalar potential
and  provide examples of dilaton dominated
supersymmetry breaking with a reasonable pattern of CP violation.

\section{Framework}

Heterotic string  models often  contain a ``hidden'' sector, i.e.
a sector which does not have direct non--gravitational interactions with the Standard
Model fields. Then it is quite plausible that supersymmetry breaking occurs in this sector
and is communicated to the visible sector by gravity. One of the popular schemes
to break supersymmetry in the hidden sector employs gaugino condensation \cite{Dine:rz}.
This possibility is quite attractive since  
a hierarchy between the Planck and SUSY breaking scales is created
dynamically  through a dimensional transmutation.
In this paper, we assume that gaugino condensation is indeed realized although
our discussion often applies more generally and is restricted by
the target space modular invariance only.

Gaugino condensation can be  realized in the $E_8 \otimes E_8$ heterotic
string theory where the condensate lives in one $E_8$, the other
forming the observable sector. After intergrating out the
condensate and any matter fields ($M$ generations transforming in
$SU(N)$) by using a truncated approximation, the
Veneziano-Yankielowicz superpotential which describes the
condensate is given by \cite{Veneziano:1982ah,Font:1990nt}\footnote{We assume
that the Kac-Moody level of the gauge group is one.}:
\begin{equation}\label{Wsimp}
  W=\tilde d \frac{e^{\frac{-3 S}{2 \tilde\beta}}}{\eta(T)^{6-\frac{9 
\delta_{GS}}{4
\pi^2\tilde\beta}}}
\end{equation}
where $\tilde\beta=\frac{3N-M}{16 \pi^2}$ is the beta function and
$\tilde d=(M/3-N)(32 \pi^2 e)^{\frac{3 (M-N)}{3N-M}}(
M/3)^{\frac{M}{3N-M}}$. 
The  K\"{a}hler potential for the dilaton and moduli is 
\cite{Witten:1985xb}:
\begin{equation}\label{K}
  K=-\ln Y-3\ln(T+\overline{T})\;,
\end{equation}
where a modular invariant combination $Y$ is given by $Y=S+\overline{S}+\frac{3}{4
\pi^2}\delta_{GS}\ln(T+\overline{T})$. The consequent scalar potential is
calculated via the supergravity relation
\begin{equation}\label{V}
  V=e^G\left( G_i \left( G^i_j \right)^{-1}G^j-3 \right)\;,
  \end{equation}
where $G=K+\ln(|W|^2)$ and the indices denote
differentiation. The sum runs over the
fields in the system ($S$ and $T$ in our case). 
Supersymmetry is broken by VEVs
of the auxiliary fields  ($j=S,T$):
\begin{equation}\label{F}
 F_j =e^{G/2}\left(G^i_j\right)^{-1} G_i\;,
\end{equation}

The superpotential describing a single gaugino condensation does not lead to dilaton 
stabilization. Thus one has to consider modifications of either the superpotential
or the K\"{a}hler potential. Some common choices are\footnote{In the context of Type I string models,
see also \cite{Abel:2000tf}.} 
(1) to employ multiple gaugino condensates \cite{multi},
(2) to postulate  S--duality \cite{Lalak:1995hn}, or (3) to incorporate non-perturbative
corrections to the K\"{a}hler potential \cite{Casas:1996zi}.
The first two options typically lead to moduli--dominated SUSY breaking 
which entails a number of  phenomenological problems
\cite{Khalil:2001dr}. 
The third  possibility is known to produce the dilaton domination (at least when
the issues of CP violation are not addressed), so we will choose this last option.
The non--perturbative   K\"{a}hler potential is assumed 
to be of the form \cite{Casas:1996zi}
\begin{equation}\label{nonperk}
  K_S=\ln \left(\frac{1}{Y}+ d \left(\frac{Y}{2}\right)^{\frac{p}{2}}e^{-b 
\sqrt{\frac{Y}{2}}}\right),
\end{equation}
where $d,p,b$ are certain constants (with $p,b >0$). This form is based
on the requirements that the non--perturbative corrections vanish in the weak
coupling limit $S \rightarrow \infty$ and that they are zero to any order 
in perturbative expansion in $1/Y$.
Dilaton stabilization with this type of the  K\"{a}hler potential has been studied
in detail in Ref.\cite{Barreiro:1997rp} with the result that
an acceptable SUSY breaking scale and  the dilaton value can be obtained, whereas
in physical cases ($K_S^S>0$) the cosmological constant does not vanish.

Minimization of the scalar potential derived from the superpotential 
 (\ref{Wsimp}) yields CP-conserving values of the modulus. To obtain CP violation,
the superpotential must be modified. In particular, it can be multiplied by
a modular invariant function $H(T)$ \cite{Cvetic:1991qm,Bailin:1998iz}:
\begin{equation}
W \rightarrow W \times H(T)\;,
\end{equation}
where 
\begin{equation}\label{H}
 H(T)=\Bigl[j(T)-1728\Bigr]^{\frac{m}{2}}j(T)^{\frac{n}{3}}P\left[j(T)\right]
\end{equation}
with $j(T)$ being the absolute modular
invariant function (see \cite{Cvetic:1991qm} for an explicit expression) 
and $P[j(T)]$ being  some polynomial. To avoid singularities in the fundamental 
domain $m$ and $n$ have to be positive integers. Although this generalized
superpotential is consistent with the modular symmetry, explicit examples
of the threshold corrections leading to this superpotential are lacking,
although the modular invariant function $j(T)$ does appear
in explicit calculations \cite{Nilles:1997vk}.

Dilaton stabilization and CP violation in models with the generalized 
superpotential were studied in Ref.\cite{Khalil:2001dr}. No phenomenologically
acceptable minima were found. However, this analysis was based on the assumptions that
$m$ and $n$ were positive. In general, this is not necessarily true   \cite{Cvetic:1991qm}
and the superpotential may have singularities at certain points in the moduli space.
In our present analysis, we allow for singularities at the fixed points of 
the modular group and find that this possibility is much more attractive phenomenologically.

Let us now list the relevant supersymmetry breaking terms.
The soft SUSY breaking lagrangian in the visible sector is

\begin{eqnarray} {\cal L}_{\rm
soft}={1\over 2}\left( M_a \lambda^a \lambda^a +{\rm h.c.} \right)
- m_\alpha^2 \hat\phi^{*\alpha}  \hat\phi^{\alpha}-\biggl(
{1\over 6}
 A_{\alpha \beta \gamma}  \hat Y_{\alpha \beta \gamma} \hat\phi^\alpha  
\hat\phi^\beta
\hat\phi^\gamma + B \hat \mu \hat H_1 \hat H_2 +{\rm h.c.}
\biggr)\;,
\end{eqnarray}
where $\hat Y_{\alpha \beta \gamma}$ and $\hat \mu$ are  the
Yukawa couplings and the $\mu$-term for the  canonically
normalized fields $\hat\phi$. With the K\"{a}hler potential and
the superpotential of the form
\begin{eqnarray}
&& K= \hat K +\tilde K_\alpha \phi^{*\alpha}  \phi^{\alpha}
+\left(  Z H_1 H_2 + {\rm h.c.} \right)\;, \nonumber\\
&&W=\hat W +  {1\over 6} Y_{\alpha \beta \gamma} \phi^\alpha
\phi^\beta \phi^\gamma  \;,
\end{eqnarray}
$\hat Y_{\alpha \beta \gamma}$ and $\hat \mu$ are given by
\cite{Brignole:1997dp}
\begin{eqnarray}
&&\hat Y_{\alpha \beta \gamma}=Y_{\alpha \beta \gamma} {\hat W^*
\over \vert \hat W \vert } e^{\hat K/2} \left( \tilde K_\alpha
\tilde K_\beta  \tilde K_\gamma \right)^{-1/2} \;,
\nonumber\\
&&\hat \mu = \left( m_{3/2} Z- \bar F^{\bar m} \partial_{\bar m}
Z\right) \left( \tilde K_{H_1} \tilde K_{H_2}\right)^{-1/2}\;.
\end{eqnarray}
Here $m=(S,T)$; $\hat{K}$ and $\hat{W}$ are the hidden sector
K\"{a}hler potential  and superpotential.
The K\"{a}hler function for a field of modular weight $n_\alpha$
is $K=(T+\overline{T})^{n_\alpha}$. For definiteness, we have
assumed the Giudice-Masiero mechanism for generating the
$\mu$-term \cite{Giudice:1988yz}. This requires the presence of a
$ZH_1H_2$ term in the K\"{a}hler potenial, which can be
implemented in even order orbifold models possessing at
least one complex structure modulus, $U$  (which we will set to
$\frac{1}{2}$). $Z$ in this case is given by \cite{Antoniadis:1994hg}
\begin{equation}
Z={1\over (T_3+T_3^*)(U_3+U_3^*)}\;,
\end{equation}

 The canonically normalized fields are
obtained by the rescaling $\hat \phi_\alpha = \tilde
K_\alpha^{1/2} \phi_\alpha$. The gaugino masses, scalar masses,
A-terms, and the B-term are expressed, respectively, as
\cite{Brignole:1997dp}:
\begin{eqnarray}
\label{softterms}
 M_a &=& {1\over 2}({\rm Re}f_a)^{-1} F^m \partial_m f_a\;,  \\
 m_\alpha^2 &=& m^2_{3/2}+V_0 - \bar F^{\bar m} F^n \partial_{\bar m} \partial_n
\ln \tilde K_\alpha\;,\nonumber\\
 A_{\alpha \beta \gamma} &=& F^m \left[ \hat K_m +\partial_m \ln Y_{\alpha \beta 
\gamma}
-\partial_m \ln (\tilde K_\alpha \tilde K_\beta \tilde K_\gamma)
\right]\;,\nonumber\\
 B &=& \hat \mu^{-1} \left(\tilde K_{H_1} \tilde K_{H_2}\right)^{-1/2} \biggl[
(2 m^2_{3/2}+V_0) Z - m_{3/2} \bar F^{\bar m} \partial_{\bar m} Z
\nonumber\\
&+& m_{3/2} F^m \left( \partial_m Z-Z ~\partial_m \ln(\tilde
K_{H_1} \tilde K_{H_2})  \right) -\bar F^{\bar m} F^n \biggl(
\partial_{\bar m} \partial_n Z - \partial_{\bar m} Z~\partial_n
\ln(\tilde K_{H_1} \tilde K_{H_2})  \biggr) \biggr]\;.\nonumber
\end{eqnarray}

Any of these terms can, in general, be complex.
However, the  EDM
measurements require them to have very small CP-phases. 
This is the notorious SUSY CP problem. 
In string models, additional difficulties arise
because $A_{\alpha\beta\gamma}$ are generically flavor--non--universal
and the flavor rotation to the basis
where the quark masses are diagonal would induce  \emph{O}(1) 
soft CP violating phases even if the soft terms were real 
initially \cite{Abel:2001cv}. The problem is exacerbated by the fact
that this rotation will produce terms proportional to the 
masses of the third generation quarks in the diagonal entries 
of the  $A$-terms.

Note that these problems arise when
SUSY and CP are broken in the same sector. That is, 
if the source of the CKM phase (in our case, complex $\langle T \rangle $)
also breaks supersymmetry. Although this is a generic situation,
we will show that this is not necessarily true and there are many vacua
in which $ F_T \sim 0$. This would remove the sources of the EDMs
due to the non--universality (the ``string'' CP problem), whereas
flavor--universal CP phases may still persist. However, the latter
are absent in our case due to the symmetry $S \rightarrow S + i\alpha$
which allows us to make the soft terms real.

\section{Patterns of the Minima}

With positive $m,n$, the minima in $T$ often fall at the fixed points of the modular group
\cite{Khalil:2001dr}. At these points the CKM phase vanishes 
\cite{Lebedev:2001qg}\footnote{If the Standard Model sector exhibited modular
invariance, this would also apply to the boundary of the fundamental domain \cite{Dent:2001cc}.
However, this is not the case in semi--realistic orbifold models \cite{Lebedev:2001qg}.}
and supersymmetry often stays unbroken. In most other cases, the minima are on the
unit circle where, again, there is no supersymmetry breaking 
($G_S=G_T=0$). A more realistic (but hard-to-achieve)
possibility is when $\langle T \rangle$ 
is inside the fundamental domain but close to the fixed point. However, this results 
in tachyons, large EDMs, and a suppressed CKM phase \cite{Khalil:2001dr}\footnote{
One may argue that observed CP violation may be mainly due to supersymmetric effects
in exotic models \cite{Brhlik:1999hs},
however such models can hardly   be motivated by strings.}.
Clearly, these vacua are not phenomenologically viable.

These problems can be rectified if we allow for negative $m$ and $n$. Indeed, this leads
to singularities at the fixed points such that the minimum is ``repelled'' from
them (since $V \sim \vert W \vert^2$) and pushed inside the fundamental domain. One should remember that
$m$ and $n$ cannot be arbitrarily large (in magnitude) negative numbers if
modulus stabilization is to be achieved. The minimum in $T$ is at a finite value 
if the superpotential diverges at $T \rightarrow \infty$ (and at its dual point, $T=0$).
At large $T$,
\begin{eqnarray}
\eta(T)^{-1} &\rightarrow& e^{\pi T/12} \;, \nonumber\\
j(T) &\rightarrow& e^{2\pi T} \;,
\end{eqnarray}
so if the polynomial $P[j(T)]$ is of degree $q$ then the divergence of the superpotential  
at infinity requires
\begin{equation}\label{condition} 
{m\over 2} +{n\over 3} > -q-{1\over 4}\;.
\end{equation}
In the pure Yang--Mills case, there are further restrictions that $H(T)$ have no
poles or zeros at infinity. This is because the asymptotic behavior of the superpotential
should match that of the threshold corrections and the simple superpotential (\ref{Wsimp}),
 i.e. $H(T)=1$, does the job \cite{Cvetic:1991qm}. This would require 
$m/2 +n/3 = -q$ with $q$ being some integer\footnote{Note that this equality cannot be satisfied 
in the conventional case $m,n>0$.}. We have analyzed supersymmetry breaking in
such cases and found that they do not lead to a reasonable phenomenology. The problem
is that the potential is often minimized at a zero of the polynomial $P[j(T)]$
where supersymmetry remains unbroken.  

\begin{table}[h]\label{table}
\begin{center}
\vspace{0.5cm}
\begin{tabular}{|c||c|c|c||}\hline
$\delta_{GS}$& 0&1&1.5\\ \hline\hline
$m$&$-\frac{1}{15}$&$-\frac{1}{30}$&$-\frac{1}{90}$\\
$n$&$-\frac{2}{15}$&$-\frac{2}{30}$&$-\frac{2}{90}$\\
$S_{\rm min}$&1.75&1.71&1.69\\
$T_{\rm min}$&$1.38+0.36i$&$1.45 + 0.44i$&$1.33+0.33i$\\
$\varphi_{_{\rm CKM}}$&${\cal O}(1)$&${\cal O}(1) $&${\cal O}(1) $\\
$V_0/M_{\rm Pl}^4$&$1.32\times10^{-32}$&$1.28\times10^{-32}$&$1.42\times10^{-32}$\\
$F_S$/Gev&-2150&-2120&-2230\\
$F_T$/Gev&$\sim$ 0&$\sim$ 0&$\sim$ 0\\
$M_a$/Gev&-604&-608&-646\\
$m_\alpha$/Gev&280&276&290\\
$A_{\alpha\beta\gamma}$/Gev&1690&1660&1750\\
$\hat \mu$/Gev&1.75&1.73&1.82\\
$\sqrt{B\hat\mu}$/Gev&280&276&290\\
\hline
\end{tabular}\\

\caption{Minima and SUSY breaking parameters.}
\end{center}
\end{table}

Inclusion of the matter fields can change the asymptotic behavior of 
the superpotential \cite{Lust:1990zi}. So, we will only require (\ref{condition}).
For simplicity we will set $P[j(T)]=1$.
In order to repel the minimum from both of the fixed points, both $m$ and $n$ should
be negative. Then, $m$ and $n$ have to be rather small in magnitude and fractional
(if $H(T)$ is to remain a rational function of $j(T)$).
Since $j(T)-1728 \sim  (T-1)^2$ in the proximity of $T=1$ and 
$j(T)\sim \left(T- e^{\pm\frac{ i \pi}{6}}\right)^3$ around
$T=e^{\pm\frac{ i \pi}{6}}$, the resulting singularities at the fixed points
are 
\begin{eqnarray}
&& H(T) \sim (T-1)^m \;\; {\rm at }\;\; T\simeq 1 \;, \nonumber \\
&& H(T) \sim \left( T-e^{\pm\frac{ i \pi}{6}}\right)^n \;\; {\rm at }\;\; T\simeq e^{\pm\frac{ i \pi}{6}}  \;.
\end{eqnarray}
We stress that since it is unclear whether or not these singularities indeed appear in explicit
models, we will take a bottom--up approach and assume $m,n$ to be free parameters
subject to the above constraints.


\begin{figure}[ht]
\begin{center}
\begin{tabular}{c}

\epsfig{file=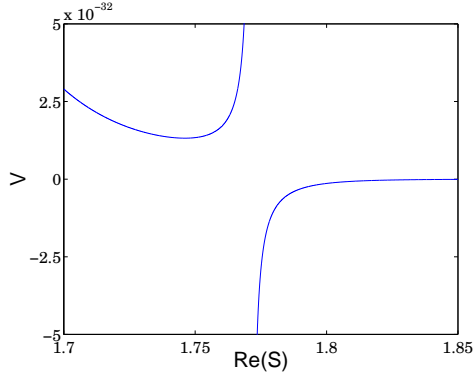, height=5cm}
\end{tabular}
\end{center}

\caption{Scalar potential with $\delta_{GS}=0$ and $m=-\frac{1}{15},
n=-\frac{2}{15}$. $T$ is set to its minimum value,
$T_{min}=1.38+0.36i$. The minimum in $S$ is at $S_{min}=1.75$. }
\label{newpaperfig2}
\end{figure}



\begin{figure}[ht]
\begin{center}
\begin{tabular}{c}

\epsfig{file=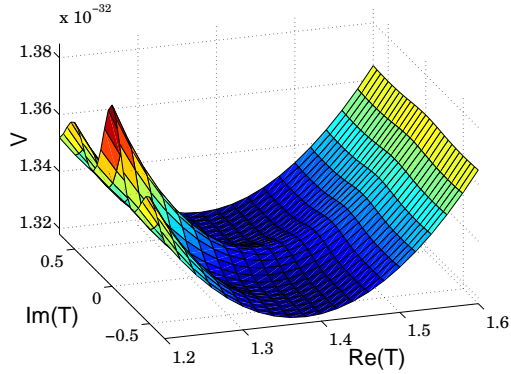, height=5cm}
\end{tabular}
\end{center}

\caption{Scalar potential with $\delta_{GS}=0$ and $m=-\frac{1}{15},
n=-\frac{2}{15}$. $S$ is set to its minimum value, $S_{min}=1.75$.
The minimum in $T$ is at $T_{min}=1.38+0.36i$. Note the invariance
of the potential under $T\rightarrow T+i$. } \label{newpaperfig1}
\end{figure}


Let us now present our numerical results.
We consider models with a single condensate and a
non-perturbative K\"{a}hler potential of the form (\ref{nonperk}).
To fix the beta function, we assume that there is one 
generation  of hidden sector matter 
in the fundamental representation of $SU(4)$.
We find that dilaton stabilization, CP violation, and reasonable
SUSY breaking can be obtained 
with, for example, $d=1, p=10, b=2$,  $\delta_{GS}=0,1,1.5$, and
$m$ and $n$ given in Table 1. 
We choose  $m$ and $n$ such that the modulus gets stabilized at
a complex value sufficiently far away from the lines Im$T= \pm 1/2$
where the CKM phase vanishes \cite{Lebedev:2001qg}. $m$ and $n$ have 
to increase with increasing $\delta_{GS}$ to produce modulus stabilization:
for $\delta_{GS}>\frac{8 \pi^2 \tilde \beta}{3}$, i.e. 1.8 in our case, 
the superpotential is no longer singular at infinity and 
$T$ does not settle at a finite value. The choice of the other parameters
is dictated by  dilaton stabilization and correct SUSY breaking scale.
The corresponding numerical results are given in Table 1.

The scalar potential for a zero $\delta_{GS}$ is shown in Figs.1 and 2.
We obtain local  minima in $S$ which are separated from the global minima by an
infinite barrier.
They always lie close to a point where $G_S^S$ vanishes and
the scalar potential diverges (Fig.1) \cite{Barreiro:1997rp}. Consequently,
at the minimum $(G^S_S)^{-1}$ is relatively large (for example, 60 in the $\delta_{GS}=0$ case).
Then, since we need the SUSY breaking scale  $F_{S(T)}$ to be around 1 TeV, in
realistic cases $m_{3/2}=e^{G/2}$ is rather small (${\cal O}$(1 GeV))  compared to $F_{S(T)}$ as seen from 
Eq.\ref{F}. The cosmological constant is in these cases positive.
Partly due to a large $(G^S_S)^{-1}$, $F_S \gg F_T$ and we have the dilaton domination. 

The Jarlskog invariant and the CKM phase can be calculated in orbifold models
assuming some  fixed point assignment to the MSSM fields.
The Yukawa coupling of the states at the fixed points $f_{1,2,3}$ belonging
to the twisted sectors $\theta_{1,2,3}$ is given by  \cite{Hamidi:1986vh},\cite{Casas:1993ac}
\begin{eqnarray}
&& Y_{f_1 f_2 f_3}=
N  \sum_{ {u} \in Z^n}
{\rm exp} \biggl[ -4\pi T \left(
{f_{23}} + {u} \right)^T
M \left( {f_{23}} + {u} \right)
\biggr]\;.
\label{yukawa}
\end{eqnarray}
where $f_{23}\equiv f_2 - f_3$, $N$ is a normalization factor, and the matrix $M$ 
(with fractional entries) is related to the internal metric of the orbifold.
A complex $T$ does not generally imply a non--zero CKM phase as the Yukawa complex
phases may be spurious and eliminated by a basis transformation. This is the case
for the prime orbifolds \cite{Lebedev:2001qg} due to restrictive 
$renormalizable$ Yukawa coupling selection rules
\footnote{This assumes that the quark fields can be associated with the fixed points
rather than their (arbitrary) combinations.}.
In the even order orbifolds, it is possible to produce the CKM phase 
at the renormalizable level with some favorable
fixed point assignment. In Table 1, we use a $Z_6-I$ example of Ref.\cite{Lebedev:2001qg}
to calculate non--removable Yukawa phases for a given $T$. In all cases they are order one.
We, however, do not address the question of the correct fermion mass hierarchy which
seems to require non--renormalizable operators \cite{fermion}.  

The models we consider possess an ``axionic'' symmetry 
\begin{equation}
S \rightarrow S + i\alpha
\end{equation}
with a real continuous $\alpha$.
Indeed, the K\"{a}hler potential is independent of Im$S$ 
and the superpotential
appears only through $\vert W \vert^2 $  such that the function $G=K+ \log\vert W \vert^2$
is invariant under $S \rightarrow S + i\alpha$.
This symmetry is a consequence of the fact that Im$S$ and Im$T$ have derivative
couplings (at least perturbatively)
\cite{Witten:dg}. The symmetry $T \rightarrow T + i\alpha$
is broken by world--sheet instanton effects down to a discrete subgroup,
while the $S \rightarrow S + i\alpha$ symmetry can only be broken by space-time
non--perturbative effects. The latter remains a symmetry of the K\"{a}hler potential
\cite{Burgess:1995aa}. 
The (approximate) invariance of the  theory under $S \rightarrow S + i\alpha$ allows
us to set $S$ and, in the case of dilaton dominated SUSY breaking, $F_S$ 
real. As a result, the soft SUSY phases are $absent$ as required by the EDM constraints.
Note, that if $F_T$ were not negligible, the axionic symmetries  would not solve
the EDM problem due to non-universality of the A-terms \cite{Abel:2001cv} and
the mechanism suggested in Ref.\cite{Choi:yd} would not work.

As seen from Table 1, the  soft breaking parameters are all of order a few 
hundred GeV except for the $\mu$-term.
The $\mu$-term is quite small due to $m_{3/2} \ll F_S$ and the fact that the Giudice-Masiero function 
$Z$ is independent of $S$. We have checked that this remains true if the non-perturbative
mechanism for generating the $\mu$-term (see e.g. \cite{mu} and \cite{Antoniadis:1994hg})   is used.
Again, the reason is that the induced $\mu$-term is of order $m_{3/2}$ which is small
compared to $F_S$. This seems to be a generic problem in such scenarios unless a different
solution to the $\mu$-problem is utilized (e.g. generating $\mu$ through a VEV of a singlet field).
It is conceivable that the $\mu$-term receives significant supergravity radiative 
corrections which may produce $\mu$ of the right size \cite{rad}.
We note that the above considered mechanisms make it difficult to achieve
radiative electroweak symmetry breaking \cite{ewsb}.

The other soft breaking parameters have reasonable values. The gaugino masses and the A-terms
are dominated by the $F_S$ contributions, while
the scalar masses and $B\mu$ receive the dominant contributions from
  $V_0$. We note that at the tree level $V_0$ coincides with the cosmological constant, whereas at the
loop level $V_0$ and the true cosmological constant receive different quadratically
divergent corrections \cite{Choi:1994xg}. Thus, a non--vanishing $V_0$ does not imply that
the cosmological constant is non--zero. In fact, in our case $V_0 \sim \tilde m^2 M_{\rm Pl}^2$
with $\tilde m$ being the typical soft breaking mass. This is exactly of the order of the quadratically
divergent 1--loop corrections  \cite{Choi:1994xg} which can be of either sign depending
on the specifics of the model. Thus, with an appropriate choice of the hidden sector it is
possible to cancel the cosmological constant.

We find that other choices of $d,p$ and $b$ give broadly
similar results and our conclusions apply quite generally.

 \section{Conclusions}

We have studied the possibility of obtaining realistic vacua in heterotic string
models possessing the $SL(2,Z)$ modular invariance. We find that dilaton stabilization,
realistic CP violation and acceptable SUSY breaking can be obtained
if (1) non--perturbative K\"{a}hler potential is used to stabilize the dilaton,
(2) the superpotential is singular at the fixed points of the modular group.

Our essential result is that it is possible to reconcile large CP violation
in the Standard Model and small CP violation in the soft SUSY breaking terms.
This necessitates dilaton dominated supersymmetry breaking and an axionic
symmetry $S \rightarrow S + i\alpha$ (which is natural in our class of models).
The only phenomenological difficulty in this case is a small tree-level $\mu$-term.
This problem may potentially be rectified by incorporating
quadratically divergent radiative corrections
\cite{rad}. The same applies to the cosmological constant.
Of course, it remains a challenge to obtain the desired properties from
explicit string models.

{\bf Acknowledgements}.
This research was supported by PPARC. We would like to thank D. Bailin for
valuable and stimulating discussions.

\end{document}